\documentclass[a4paper,11pt]{article}
\usepackage{pos}
\usepackage{graphicx}
\usepackage{caption}    
\usepackage{adjustbox}  
\usepackage{siunitx}

\title{Cosmic ray detection with the LOFAR radio telescope}

\author*[a]{K.~Terveer}
\author[a]{S.~Bouma}
\author[b,c]{S.~Buitink}
\author[b,c]{A. Corstanje}
\author[c]{M. Desmet}
\author[b,d,e]{H. Falcke}
\author[e]{B.M. Hare}
\author[b,d,a]{J.R. Hörandel}
\author[f,a]{T. Huege}
\author[f]{N. Karastathis}
\author[a]{P. Laub}
\author[b,d]{K.~Mulrey}
\author[g,a]{A. Nelles}
\author[h,i]{O. Scholten}
\author[e,h]{P. Turekova}
\author[j]{S. Thoudam}
\author[k]{G. Trinh}
\author[e]{S. ter Veen}

\affiliation[a]{Erlangen Centre for Astroparticle Physics, Friedrich-Alexander-University Erlangen-Nürnberg, 91058 Erlangen, Germany}
\affiliation[b]{Department of Astrophysics/IMAPP, Radboud University Nijmegen, P.O. Box 9010, 6500 GL Nijmegen, The Netherlands}
\affiliation[c]{Vrĳe Universiteit Brussel, Astrophysical Institute, Pleinlaan 2, 1050 Brussels, Belgium}
\affiliation[d]{Nikhef, Science Park Amsterdam, 1098 XG Amsterdam, The Netherlands}
\affiliation[e]{Netherlands Institute for Radio Astronomy (ASTRON), Postbus 2, 7990 AA Dwingeloo, The Netherlands}
\affiliation[f]{Institut für Astroteilchenphysik, Karlsruhe Institute of Technology (KIT), P.O. Box 3640, 76021 Karlsruhe, Germany}
\affiliation[g]{Deutsches Elektronen-Synchrotron DESY, Platanenallee 6, 15738 Zeuthen, Germany}
\affiliation[h]{University of Groningen, Kapteyn Astronomical Institute, Groningen, 9747 AD, Netherlands}
\affiliation[i]{Interuniversity Institute for High-Energy, Vrije Universiteit Brussel, Pleinlaan 2, 1050 Brussels, Belgium}
\affiliation[j]{Department of Physics, Khalifa University, P.O. Box 127788, Abu Dhabi, United Arab Emirates}
\affiliation[k]{Department of Physics, School of Education, Can Tho University Campus II, 3/2 Street, Ninh Kieu District, Can Tho City, Vietnam}

\emailAdd{karen.terveer@fau.de}

\abstract{The LOw Frequency ARray (LOFAR) has successfully measured cosmic rays for over a decade now. With its dense core of antenna fields in the Netherlands, it is an ideal tool for studying the radio emission from extensive air showers in the \SI{e16}{\electronvolt} to $10^{18.5}$ \si{\electronvolt} range. Every air shower is measured with a small particle detector array and hundreds of antennas, which sets LOFAR apart from other air shower arrays.
We present our current achievements and progress in reconstruction, interpolation, and software development during the final phases of measurement of LOFAR 1.0, before the LOFAR array gets a significant upgrade, including also plans for the final data release and refined analyses.}

\FullConference{10th International Workshop on Acoustic and Radio EeV Neutrino Detection Activities (ARENA2024)\\
11-14 June 2024\\
The Kavli Institute for Cosmological Physics, Chicago, IL, USA\\}

\begin{document}
\maketitle

\section{LOFAR}
The LOw Frequency ARray (LOFAR) is the worlds largest radio telescope, consisting of numerous antenna fields (stations) distributed across Europe. There are 52 stations in total, 24 of which form a dense core in the Netherlands \cite{generalLOFAR}. It is a shared telescope that is primarily used for astronomy observations.

For cosmic ray detection the core stations are utilized, each equipped with 96 Low Band Antennas (LBA) that consist of two perpendicularly arranged dipole antennas operating in the 30-\SI{80}{MHz} band. In addition, the core is equipped with 20 scintillator units (LORA) which are used for triggering and auxiliary air shower reconstruction \cite{LORA}. Once LORA detects muons from a cosmic ray induced air shower, the buffered data from the antennas are read out.

LOFAR measures the radio emission stemming from the dominant geomagnetic effect and the sub-dominant Askaryan effect. In the LOFAR frequency range (\SI{30}{\mega\hertz} - \SI{80}{\mega\hertz}) the two emission mechanisms lead to the characteristic asymmetric bean-shaped footprint on the ground, which the antennas sample and which can be used for reconstructing shower properties such as the shower maximum $X_\text{max}$ or the cosmic-ray energy. At the LOFAR site, the radio footprint of a shower can range between a few hundred meters to a kilometer in diameter, hence the dense instrumentation of antennas plays out its power when studying the features of the footprint.

The LOFAR-method for reconstruction of the $X_\text{max}$ has established itself as the current state-of-the art, relying on a series of simulations for each measured shower and performing a $\chi^2$-minimization, yielding a systematic uncertainty of 7-\SI{9}{\gram\per\centi\meter\squared} \cite{xmax}. The method was also used by the Pierre Auger Observatory \cite{xmaxauger}.

LOFAR targets an energy range of \SI{e16}{\electronvolt} to $10^{18.5}$ \si{\electronvolt} which is precisely the expected transition region from galactic to extragalactic sources.

\section{Current developments}
Recent developments in the LOFAR Key Science Project Cosmic Rays involve, among others, the switch to a new reconstruction software, the usage of interpolation of air shower simulations and new, simulation-independent approaches to reconstruction. In this section we will highlight some of the work done in the past year.

\subsection{Software}
A new standardized software to use for all analyses has been in development for a while, motivated by no longer maintainable custom reconstruction solutions on continuously upgraded server infrastructure. Instead of developing something stand-alone for cosmic ray detection, there are significant advantages of joining an already existing software used by other radio experiments. This ultimately led the choice of NuRadioReco \cite{NRR}, a python framework for detector simulation and reconstruction of radio detectors for neutrinos and cosmic rays.
Advantages of NRR include its modular structure, existing documentation and active developers as well as the many pre-existing modules that can be conveniently used once a detector description exists. 
In the past year we have integrated the full LOFAR reconstruction pipeline \cite{pipeline} into NuRadioReco: This required both the full detector description, geometry and antenna and hardware responses, and all modules that are related to data processing. The latter includes the read-in of events and the subsequent pulse-finding, RFI filtering, Galactic calibration and direction fitting of the data.

\begin{figure}
    \centering
    \includegraphics[width=0.99\linewidth]{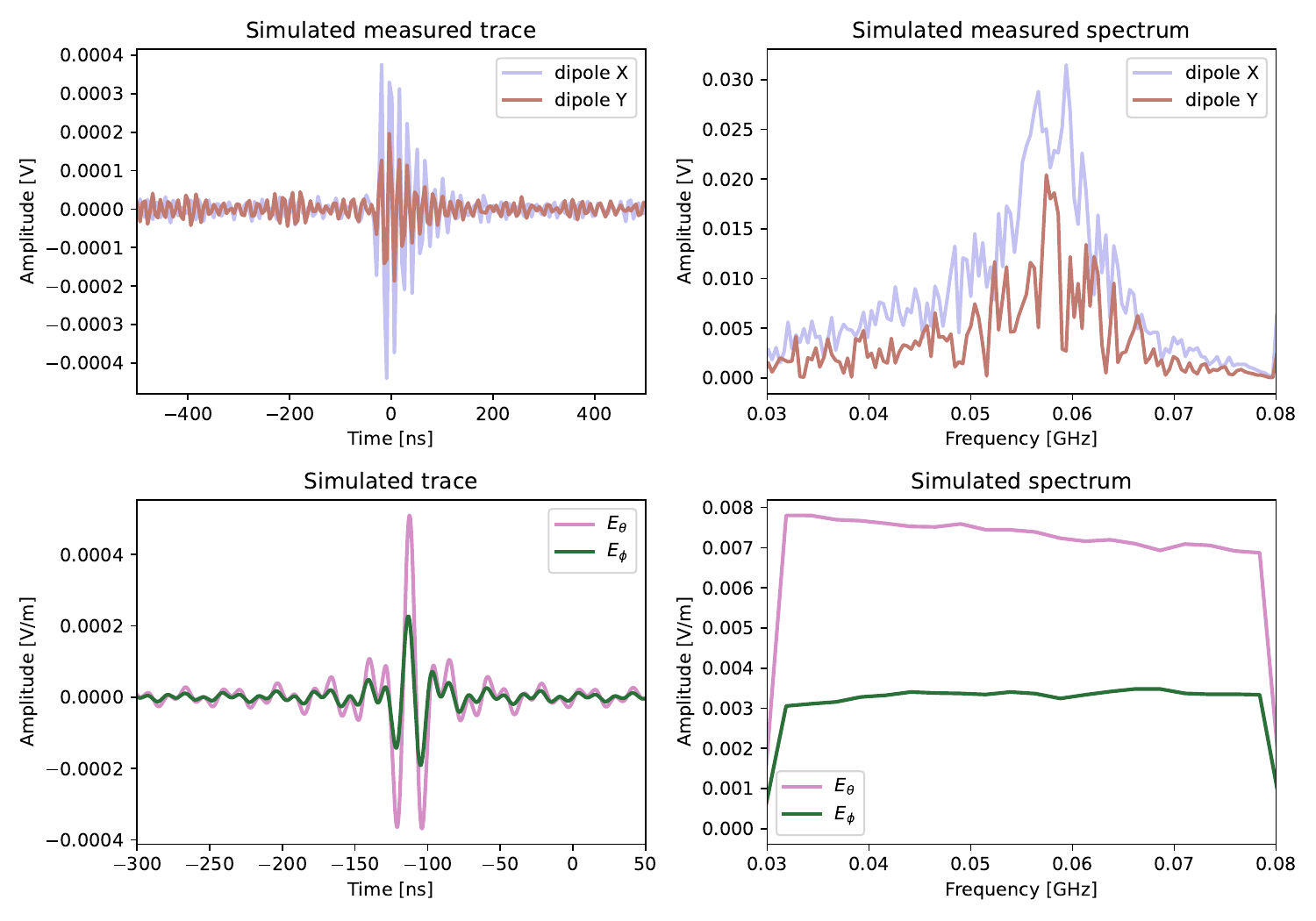}
    \caption{\textit{Top}: Simulated trace forwarded through the LOFAR signal chain (antenna and hardware) in NuRadioReco, with added Galaxy noise. \textit{Bottom}: Original Electric field trace from CoREAS star-shape simulation fed into NuRadioReco and interpolated to antenna location with the interpolator described in \ref{interpolation}.}
    \label{fig:NRR}
\end{figure}

In addition to reading in data for analysis, we can now also create realistic data from simulations by applying our signal chain to simulated pulses and adding noise. One example of that can be seen in Figure \ref{fig:NRR}, which shows a simulated "measured trace" and the corresponding simulation truth. This will enable us to test our analyses on synthetic data for verification. This full circle, i.e. simulating and reconstructing with the same software had never been available for LOFAR data before. 

\begin{figure}
    \centering
    \begin{minipage}[b]{0.65\textwidth}
        \adjustbox{valign=b}{\includegraphics[width=\linewidth]{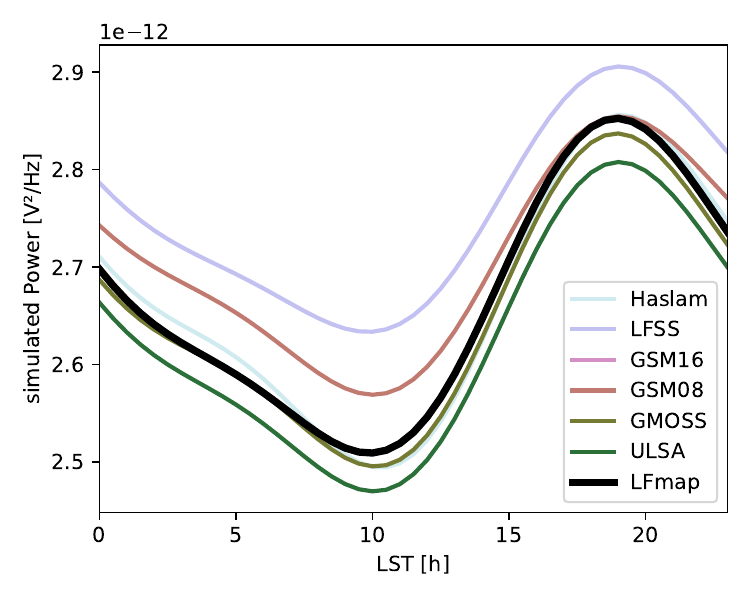}}
    \end{minipage}
    \hfill%
    \begin{minipage}[b]{0.34\textwidth}
        \captionof{figure}{Simulated average integrated power of the Galaxy for one LOFAR antenna over the sidereal day as predicted by different sky models. The sky model used in the previous absolute calibration, LFmap, is marked as a bold black line.}
        \label{fig:NRR_galaxy}
    \end{minipage}%
\end{figure}

Not only do we use existing modules, but we ourselves are also contributing to NuRadioReco by adding our own analyses, modules and improvements to code. One example is the module that simulates the radio emission from the Galaxy, to which we added various sky maps that all users can choose from. The sky models had been compiled in \cite{skymodels}. They rely on various reference measurements at different latitudes and frequencies, which makes some sky models more suitable for a given experiment than others. We will make use of this not only when creating synthetic data, but also for the absolute calibration of the antennas \cite{calibration} which we are currently reviewing in the process of finding a possible cause for the tension between the mass composition reported by LOFAR and AERA \cite{xmax}. As the energy reconstruction is dependent on the fluence of the signals which in turn is dependent on the absolute antenna calibration, a shift in calibration might cause a shift in reconstructed energy which could affect the compatibility of LOFAR and AERA. In Figure \ref{fig:NRR_galaxy} we observe that for different sky maps the integrated power differs slightly from the sky map that was originally used in the calibration (LFmap, \cite{lfmap}). Here, the integrated power was calculated for each LST via

\begin{equation}
    P=\int_{\SI{30}{\mega\hertz}}^{\SI{80}{\mega\hertz}}|F(\nu)| d\nu ,
\end{equation}

where $F(\nu)$ is the frequency component of the galaxy noise after propagating through the signal chain. A Fourier series was fit to the power values, as done in the original calibration. It is yet to be determined what impact the choice of sky model has on the calibration, however so far it seems the variation introduced by the different sky models alone is not large enough to account for the needed energy shift. This however is still under active investigation, along with the question of which the best sky model for the calibration would be. In any case, a completely new re-implementation of the calibration will be a good cross-check for potential issues. 

Another contribution we made to NuRadioReco was implementing a new CoREAS interpolator, described in the following subsection. This interpolator can in the future be used by all users of NuRadioReco and significantly reduces the computational demand of air shower simulations for large arrays with many antennas while retaining the needed accuracy.

\subsection{Interpolation}\label{interpolation}
With growing detector arrays, air shower simulations with software such as CoREAS \cite{CoREAS} get more and more computationally intensive. For a LOFAR sized detector, one shower simulation takes up 1 to 3 days of CPU time with a dedicated grid that allows a simple interpolation of the scales of LOFAR. With regard to next generation array such as the Square Kilometre Array (SKA) \cite{SKA} we are looking at arrays the size of 60,000 antennas. For the LOFAR-method of reconstructing the $X_\text{max}$, around 30 simulations are run for each detected shower. In the SKA era, this would become too cost and energy intensive. With the interpolation algorithm we present in \cite{arthur_interpolator}, it suffices to simulate a shower for approximately 200 antennas while retaining enough accuracy for high-precision analysis. Also, previous LOFAR analyses have only used the signal fluence and its interpolated footprint to determine shower parameters. With the advent of SKA, one would like to use a full waveform interpolation (e.g.\ for interferometery), which is also part of the implemented interpolator. 

\begin{figure}
    \centering
    \includegraphics[width=0.99\linewidth]{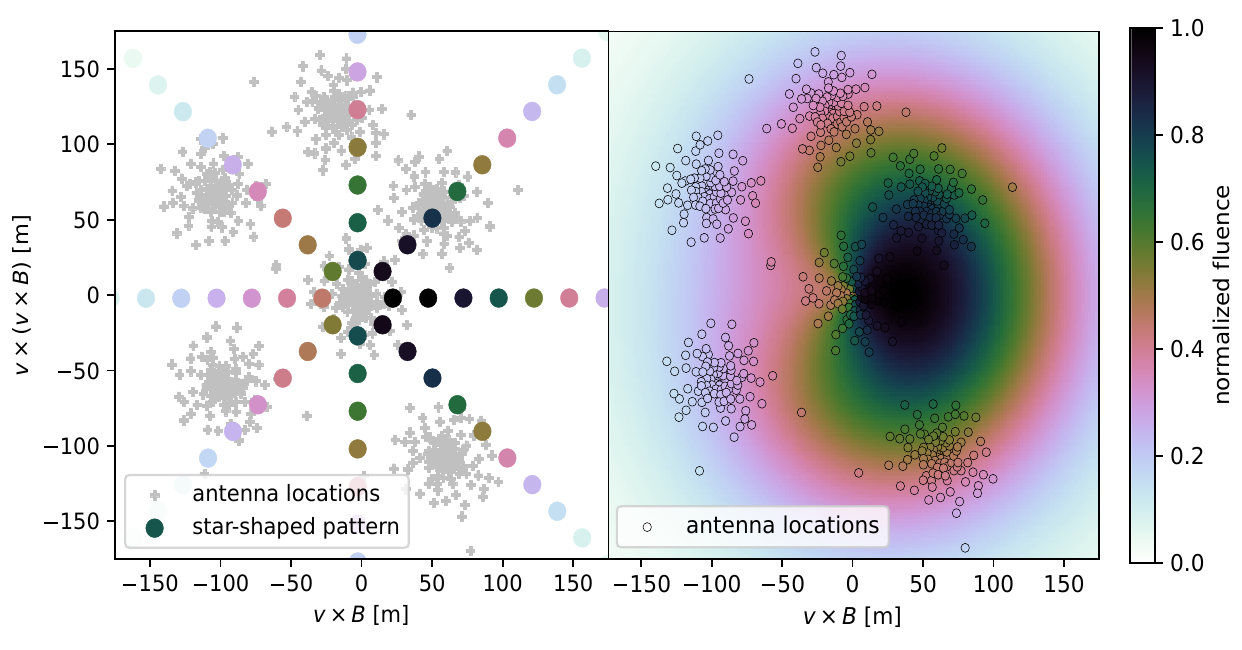}
    \caption{\textit{Left}: the simulated star-shaped pattern, the color indicating the normalized fluence at that position, and the grey crosses in the background representing the LOFAR antenna positions that the simulation will be interpolated to. \textit{Right}: The fluence interpolation shown as gradient in the background and the corresponding fluence values calculated from the interpolated pulses shown for each antenna location. Each antenna dot is filled with the to the fluence corresponding color.}
    \label{fig:interpolation}
\end{figure}

The interpolation method is based on Fourier series and cubic splines. After simulating a radio shower with CoREAS for a star shaped antenna array, we use an FFT along a circle at each radius $r$ from the shower core, to obtain Fourier series components for the fluence:
\begin{equation}
\hat{f}(r, \theta)=\sum_{k=0}^{n / 2} c_k(r) \cos (k \theta)+s_k(r) \sin (k \theta)
\end{equation}
These coefficients are then radially interpolated using cubic splines, allowing to get the fluence at any desired antenna position. The time series is modeled as an amplitude spectrum with a phase:
\begin{equation}
F(v)=\mathcal{F}(E(t))=|F(v)| \exp (\mathrm{i} \phi(v)),
\end{equation}
and by applying the same interpolation method to the spectral amplitudes and phase factors, the full pulse time series can be obtained at any position.

Compared to the previous method of interpolation based on radial basis functions, with this method the fluence interpolation error is reduced by a factor of above 3 in the LOFAR frequency range. Figure \ref{fig:interpolation} shows an exemplary result of using the interpolator as implemented in NuRadioReco, showing the simulated star-shape antennas, the interpolated LOFAR antennas of six stations and the full interpolated radio footprint. Additionally, Figure \ref{fig:NRR} (bottom) shows interpolated pulses and their frequency spectrum for one antenna location. For the 30 to 80 MHz band, the typical error is only \SI{0.3}{\percent}, while the timing error is around \SI{0.04}{\nano\second}.

\subsection{Machine learning based reconstruction}
With the current standard reconstruction being so heavily based on simulations, we have explored the possibility of using machine learning instead. For that, a model was trained on the already existing 300,000 air shower simulations with a simple noise model added to it. 

\begin{figure}
    \centering
    \begin{minipage}[b]{0.67\textwidth}
        \adjustbox{valign=b}{\includegraphics[width=\linewidth]{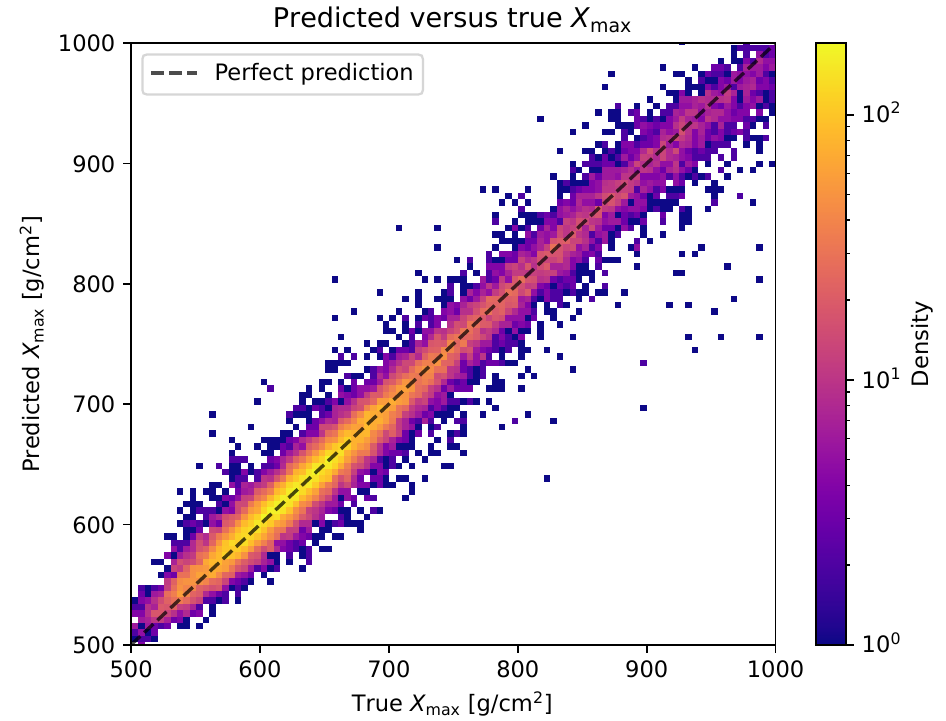}}
    \end{minipage}
    \hfill%
    \begin{minipage}[b]{0.31\textwidth}
        \captionof{figure}{Reconstruction results for the $X_\text{max}$ value, showing the predicted values from the model versus the MC truth for each simulation. For most simulated cosmic rays, the values align nicely around the perfect prediction axis.}
        \label{fig:machine_learning}
    \end{minipage}%
\end{figure}

First, the radio pulse was characterized by features: Peak time, total power (integral over the Hilbert envelope of the time trace), the parameters of a quadratic fit performed on the Amplitude spectrum in the frequency domain, and the atmosphere information of the simulation. The model, a Fourier Convolution Neural Network (FCNN), was then trained on these features to predict targets: $X_\text{max}$, $L$, $R$, $E$ and the core position.
$R$ and $L$ are quantities that relate to the rise and fall of the lateral shower distribution, they can be introduced in a reparametrisation of the better known Gaisser-Hillas formula, the $R,L$ formula \cite{RL}:
\begin{equation}
N_c^{\mathrm{R}-\mathrm{L}}\left(X_{\mathrm{z}}\right)=N_{\max } \times\left(1-\frac{R}{L}\left(X_{\max }-X_{\mathrm{z}}\right)\right)^{R^{-2}} e^{\frac{X_{\max }-X_{\mathrm{z}}}{R L}}
\end{equation}
The FCNN contains seven hidden layers, four million trainable parameters and was given 3 hours of training time. It yielded a mean absolute error on $X_\text{max}$ of \SI{12.2}{\gram\per\centi\meter\squared} or \SI{2.4}{\percent}, which is vizualised in Fig. \ref{fig:machine_learning}. For the energy reconstruction, the MAE of the $\log{E}$ was 0.0305, the shower core was resolved within \SI{3.3}{m}$\pm$\SI{2.4}{m} while $L$ and $R$ were the least accurate, with L yielding an MAE of \SI{6.31}{\gram\per\centi\meter\squared} (\SI{13}{\percent}) and for $R$ the MAE was 0.022 or \SI{11}{\percent}. Caution needs to be taken when interpreting the MAE for the primary energy, as the model learned a correlation between atmosphere information and the energy from the simulations it was trained on. This is something to look into in future extensions of this work.

The MAE values suggest that there was not enough information on $L$ and $R$ in the features that are given as inputs to the model. As first attempt on a ML based reconstruction, these results are, however, promising. In the future, this work could be extended by adding realistic noise to the simulation data and investigating the changes in performance. Additionally, the reconstruction of $X_\text{max}$, $L$ and $R$ would benefit strongly from including data from the High Band Antennas (HBA), which currently are challenging to use for air shower detection due to the data being already beamformed upon read-out. This will change in the next phase of the telescope, LOFAR2.0, which is currently at the start of being implemented, as discussed in \cite{proceedingsKatie}. The full ML analysis can be found in \cite{LuukThesis}.

\subsection{Holistic Air shower reconstruction}
The ideal reconstruction algorithm would be a physics-informed one that does not rely on unique simulations to compare to for each event. Additionally, the classical methods of reconstruction (e.g.\ the LOFAR-method) do not use all information that is available in our data -- for the $X_\text{max}$ reconstruction, only the footprint shape is used, while for the energy reconstruction \cite{energyReco} the fluence is used. Neither method uses the pulse shape or the wavefront timing. 

We have started efforts towards a holistic reconstruction that takes in all available information from the data: timing, polarization, signal strength, footprint, and also in addition the data available from the LORA particle detectors. This poses a complex challenge of ensuring the model is physics informed, i.e.\ finding a parametrization that relates measured quantities to physical ones such as $X_\text{max}$, $L$, $R$ or the energy, but leaving enough room for deviations from the parametrization due to unknowns in the data and imperfect parametrizations.
It also requires a modeling of the noise. A method that has been well tested in the field of astronomy is a promising candidate for these prerequisites: Information Field Theory \cite{IFT} can be used to reconstruct signals from noisy data given prior information. It has in the past already been successfully used for reconstructing radio pulses for the use cases of both in-ice detection and air showers \cite{IFTRadio}, which we are currently working on applying to LOFAR as a first reconstruction step.

A further advantage of this method will be that it could be used on all candidate events, even those with low signal-to-noise ratio (SNR), while simultaneously providing reliable uncertainties on all reconstructed parameters. 

\section{Outlook}
After more than a decade of continuous observation, the first phase of the Low Frequency Array, LOFAR1.0 is coming to an end, and observation with the core stations is temporarily interrupted while the hardware upgrades for the next phase, LOFAR2.0, are installed \cite{proceedingsKatie}. During this downtime, we will (re-)analyse the full set of LOFAR1.0 data and expect to subsequently release the results. The work that has been done on software in the previous year serves as the ideal basis for the upcoming analyses. This will also be the opportunity to test new reconstruction approaches along with the standard method, which will all serve as preparation for following observations with LOFAR2.0 and also SKA-low.

\end{document}